\documentclass[reprint,aps,prl,showpacs]{revtex4-2}

\usepackage[utf8]{inputenc}
\usepackage{graphicx}
\usepackage{tikz}
\usepackage{dcolumn}
\usepackage{amsmath}
\usepackage{amsfonts}
\usepackage{amssymb}
\usepackage{hyperref}
\usepackage{pbox}
\usepackage{pinlabel}
\usepackage{ragged2e}
\usepackage[justification=RaggedRight]{caption}
\usepackage[singlelinecheck=off]{subcaption}
\usepackage{ulem}

\newcommand\djf[1]{\ {\color{teal}{SF: #1}}\ }

\begin{document}

\title{Strange metal transport from coupling to  fluctuating spins}


\author{
	Simone Fratini$^{1\ast}$,
    Ivan Duchemin$^{2}$,
	Arnaud Ralko$^{1}$,
	Sergio Ciuchi$^{3}$ \\
	\small$^{1}$Universit\'e Grenoble Alpes, CNRS, Grenoble INP, Institut Ne\'el, 38000 Grenoble, France. \\
	\small$^{2}$Universit\'e Grenoble Alpes, CEA, IRIG-MEM-L\_Sim, 38054 Grenoble, France \\
	\small$^{3}$Dipartimento di Scienze Fisiche e Chimiche, Universit\'a dell’Aquila, Coppito-L’Aquila, Italy. \\
	\small$^\ast$Corresponding author. Email: simone.fratini@neel.cnrs.fr 
}

\begin{abstract}

Metals hosting strong electronic interactions, including high-temperature superconductors, behave in ways that do not conform to  normal Fermi liquid theory. 
To pinpoint the microscopic origin of this strange metal behavior, here we reexamine the 
d.c. and frequency-dependent conductivity
of the two-dimensional t-J model taking advantage of recent improvements made on the finite temperature Lanczos method, enabling numerically exact  calculations at unprecedentedly low temperatures and high spectral resolution. We find that strange metallicity is pervasive in the temperature-doping phase diagram whenever anti-ferromagnetic order is suppressed, and advocate that key insights on Planckian relaxation can be gained by extending the study to the frequency and time domain. Our results indicate that Planckian behavior 
does not 
originate from the scattering properties of the current carriers,
being  instead rooted in the quantum statistical nature of the charge response.

\end{abstract}

\maketitle


Strongly correlated materials often exhibit resistivities that increase linearly with temperature from few to hundreds of degrees K, at odds with the standard theory of metals \cite{PhillipsScience22}. These anomalies are part of a broader, puzzling picture, as they also extend to the frequency domain: associated with the strange behavior of  charge transport, the decay of the conductivity with frequency is slower than that associated with 
exponential relaxation processes, 
and therefore incompatible with normal diffusion at long times \cite{BasovRMP}.

Despite broad experimental evidence of the phenomenon, accurate numerical calculations of strange metal transport in correlated electron systems have only appeared  very recently, and there is no consensus on its precise microscopic origin. These numerical studies have mostly focused on the Hubbard model \cite{DengPRL13,PakhiraPRB2015,Kokalj,VucicevicPRL19,Devereaux,WangDevereaux25,Kovacevic-PRL25,Stepanov-PRL26} and its strong coupling analog, the t-J model \cite{JaklicPRB95,Prelovsek-PRL95,TsunetsuguJPSJ97}, due to their direct microscopic connection with the experimental systems. Alternatively, more phenomenonology-driven descriptions have been proposed enforcing quantum criticality \cite{CapraraCommPhys22} and the absence of quasiparticles, 
\cite{PatelScience23,PatelPRX25}, for their ability to capture some of the observed physical behavior. 
Important insights have been gained from local theories \cite{DengPRL13,PakhiraPRB2015}, and from the study of the quasiparticle scattering rate alone \cite{TremblayPNAS22,CapraraCommPhys22}. However, when dealing with an unknown transport mechanism  it is essential to undertake the much harder task of fully calculating the transport observables:  these contain non-local and quantum interference effects
--- so-called vertex corrections \cite{VucicevicPRL19,TsunetsuguPRB16,Millis-PRB22,Kovacevic-PRL25,Stepanov-PRL26,RammalPRL24} --- as well as more fundamental properties of response functions that are not included in the one-particle properties and that, as we show here, 
play a key role in the observed behavior. 



We choose to work with the t-J model. Being equivalent to the Hubbard model in the large-$U$ limit, the t-J model has a direct connection with the materials' microscopics,
retaining the crucial anti-ferromagnetic interactions $J= 4 t^2/U$ arising from particle exchange at finite $U$. To our advantage, however,
the strength of anti-ferromagnetic correlations can be tuned independently of the correlation strength, which makes it an ideal framework to disentangle the role of spin fluctuations from the many-body effects related to the charge. 


Our results comprehensively demonstrate strange metal behavior arising from the coupling of the charge carriers with spin fluctuations.
This is characterized by $T$-linear dependence of the resistivity and a finite $T=0$ residual resistivity, suggestive of residual spin disorder persisting down to the lowest temperatures.
Notably, the conductivity appears to scale proportionally to the number of available holes in a broad range of dopings, $0<p\lesssim 0.25$, suggesting that the strange carriers behave independently from one another.    
By extending our analysis to the frequency domain, we identify that Planckian relaxation does not originate from the temperature dependence of the carrier scattering properties. It is instead a consequence of the large scattering rate that characterizes bad metals to start with: 
when this is larger than the thermal scale $k_BT/\hbar$, the low-frequency conductivity and the resistivity become dominated by the quantum statistical properties of the charge response. In addition to explaining the non-Drude shapes observed in experiments, this finding implies that the emergence of a narrow low-frequency peak in the optical conductivity of bad metals
should not 
be seen as the signature of a separate conduction channel \cite{Kumar,vanHeumen22}, but rather as the quantum 
response of the same type of bad metallic carriers.




\textit{Model and method.---}
The t-J model Hamiltonian is $H=-\sum_{ijs}t_{ij }(\tilde c^\dagger_{is} \tilde c_{js} +h.c.) + J_{ij } (\vec{S}_i\cdot \vec{S}_j-\frac{1}{4}n_in_j)$, where $\tilde c^\dagger_{is}$ and $\tilde c_{js}$ are projected creation and annihilation operators for electrons on sites $i$ and $j$, that take into account  the constraint of no local double occupancy enforced by the strong Coulomb repulsion limit, $n_i$ is the density at site $i$ and $J_{ij}$ is an antiferromagnetic (AF) exchange coupling between the corresponding quantum electron spins $\vec{S}_i$ and $\vec{S}_j$.
We focus on the square lattice, setting $t_{ij}=t$ and $J_{ij}=J$ for nearest neighbors and  taking $t$ as the energy unit. We also consider frustrated cases with  $t_{ij}=t^\prime$ and $J_{ij}=J^\prime$ for next-nearest neighbors. 

We solve the t-J model numerically using the finite-temperature Lanczos method (FTLM) \cite{prelovsek} as well as its low-temperature generalization (LTLM) \cite{Aichhorn03}  on clusters of size up to $N_s=24$ sites, supplemented by twisted boundary condition (TBC) averaging to increase the effective cluster size (see methods and supplement). Additionally, we generalize and improve a recently benchmarked prescription for a reliable extrapolation of the Kubo formula to the d.c. limit \cite{RammalPRL24}.  Taken together, this combined methodology allows to study the strange metal phase reaching high spectral resolution and unprecedentedly low temperatures, reaching a lower limit of $T/t\simeq 0.01$. When converted to cuprate units ($t\simeq 0.4$eV) this corresponds to $T\approx 40$K.


\textit{Strange metal behavior of a single hole.---}
\begin{figure}
    \centering
    \includegraphics[width=9cm]{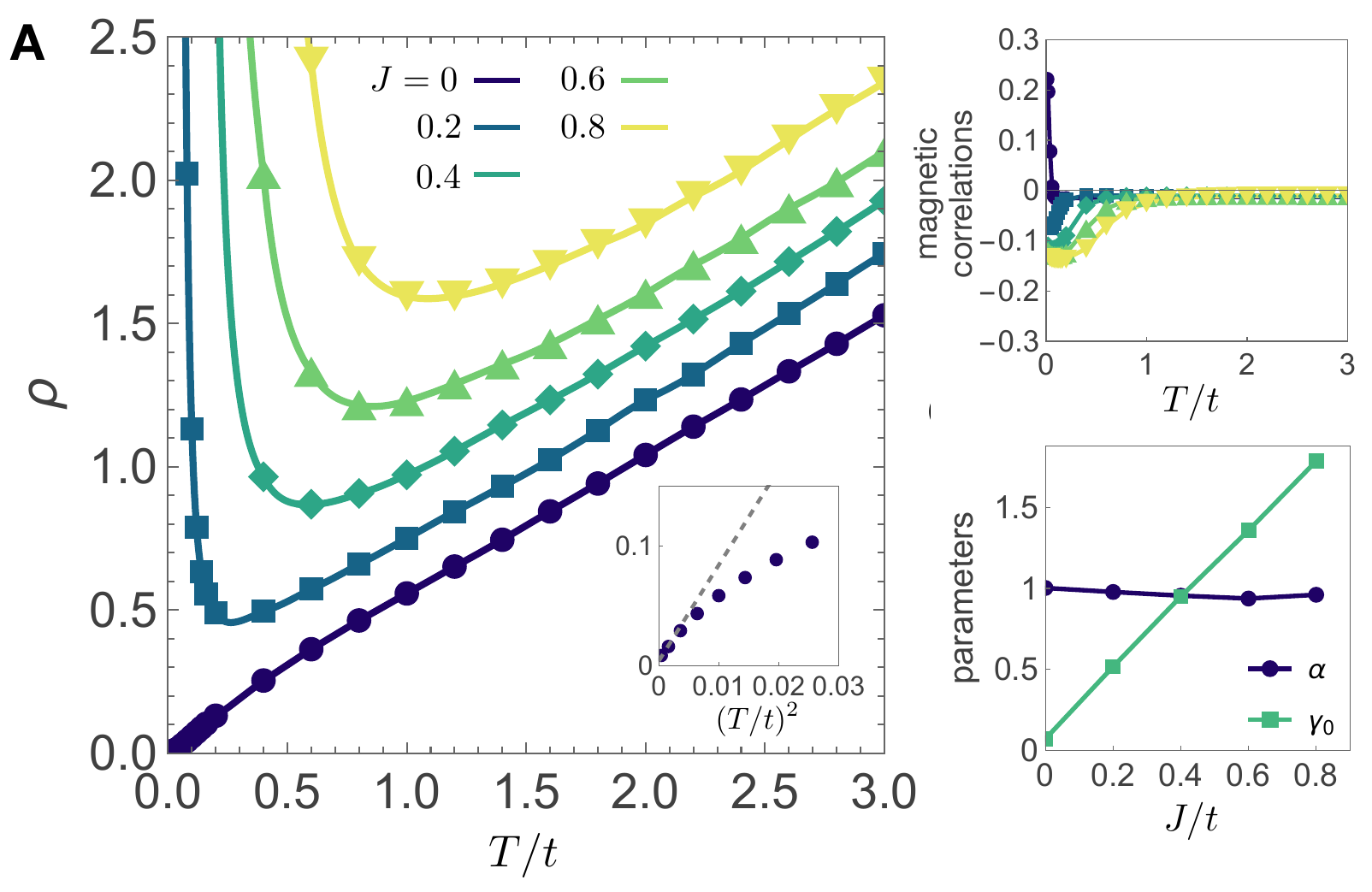}
    \caption{
   \textbf{Strange metal behavior in the t-J model.} 
   (\textbf{A}) Resistivity per hole (inverse mobility) obtained from FTLM at different values of the AF exchange coupling (labels). 
   Simulation parameters are $N_s=18$, $M=400$ Lanczos iterations and $N_\phi=49$ twisted boundary conditions (for $J=0$ we use the low-temperature Lanczos method and a larger size $N_s=24$, $M=800$, $N_\phi=105$ in order to access lower temperatures). 
   For layered two-dimensional materials with interlayer separation $d$, physical units are restored upon multiplying the dimensionless resistivity by $\bar{\rho}=\hbar/e^2 (d/x)$ for a concentration $x$ of independent holes. 
   (\textbf{B}) Spin correlations $\langle \hat n_0 \ \vec{S}_i\cdot \vec{S}_j \rangle$ between a site $i$ adjacent to the hole  and $j$ at the maximum allowed distance on the opposite sublattice (hole at site $0$). Antiferromagnetic correlations arise at $T\lesssim J$ except for $J=0$, where Nagaoka ferromagnetism is established at very low temperatures \cite{Park}. 
    (\textbf{C})  Slope and $T=0$-intercept of the apparent transport scattering rate, Eq. (\ref{Drude-analysis}). The inset of panel (\textbf{A}) shows $\tilde\rho$ vs $T^2$  for $J=0$.
   }
    \label{fig:Fig1}
\end{figure}
We first present our results for one hole in an unfrustrated square lattice, $t^\prime = 0$, in order to understand the effects of the magnetic background. 
Fig. \ref{fig:Fig1}A reports the dimensionless resistivity per hole, $\tilde{\rho}$, calculated for different values of the antiferromagnetic exchange interaction $J$. 
Remarkably, the resistivity in the \textit{paramagnetic} phase at $J=0$ is linear in $T$ with a constant slope all the way from high temperature down to $T_{\mathrm{FM}}\simeq 0.1t$, marking the onset of Nagaoka ferromagnetism \cite{Park}. 
Below this temperature, normal $\rho \propto T^2$ behavior is recovered (inset of Fig. \ref{fig:Fig1}A and Fig. \ref{fig:Fig2}F, blue solid line). 


Although, as we show below, this does not apply to strange metals \cite{ZaanenSciPost18}, we proceed to analyze the calculated resistivity  following the same procedure that is customarily used in experiments  \cite{Bruin,Legros}, i.e. using the Drude model, $\sigma_D=ne^2\tau_{\mathrm{tr}}/m$, $\rho_D=1/\sigma_D$. In this framework, with the band mass $m$ and the carrier density $n$ given, the transport properties 
are entirely encoded in the transport scattering time $\tau_{\mathrm{tr}}$. Strange metals are \textit{defined} by the observed linear relation 
\begin{equation}
   \hbar/\tau_{\mathrm{tr}}= \alpha k_BT + \gamma_0,
   \label{Drude-analysis}
\end{equation}
with $\gamma_0$ a constant.
When the  numerical prefactor $\alpha$ is  $1$,  the relaxation is termed Planckian \cite{Zaanen04,Zaanen18,Bruin,Legros}.

Using $m=\hbar^2/(2ta^2)$ for a square lattice with lattice spacing $a$, and taking the carrier density $n$ equal to the hole density $p$, we can extract the apparent transport scattering time from the data as $\hbar/\tau_{\mathrm{tr}}=2 t \tilde{\rho}$. 
The linear slope extracted from Fig. \ref{fig:Fig1}A for $J=0$ is $\alpha=1.003$ (see panel C). This is the first important result of our study: 
At low hole concentrations,
the metallic state of the infinite-$U$ Hubbard model, represented here by the t-J model for $J=0$,  is a Planckian metal at all temperatures above the ferromagnetic transition. 
Because charge fluctuations are absent in this limit,  the cause of the observed Planckian behavior is necessarily the presence of the spin fluctuating background, that is in this case paramagnetic and thermal.

The effect of a finite antiferromagnetic (AF) exchange coupling, $J>0$, is also depicted in Fig. \ref{fig:Fig1}A. AF correlations build up upon lowering the temperature (shown in Fig. \ref{fig:Fig1}B) until the system orders anti-ferromagnetically at $T=0$ \cite{LugasPRB06,JiangPNAS21,XuPRR22}, 
resulting in insulating behavior as $T\to 0$. 
A $T$-linear resistivity is recovered at all $T\gtrsim J$, i.e. as soon as the AF correlations vanish  (Fig. \ref{fig:Fig1}B), extrapolating to a finite residual resistivity $\rho(T=0) \propto J$. The recovery of uncorrelated spin dynamics is also indicated by the entropy approaching the value $k_B \log 2$ per spin  (see supplement). 
From the Drude analysis, we obtain the same Planckian slope $\alpha \simeq 1$  for all finite $J$ and an apparent residual scattering rate  $\gamma_0 \approx 2J$ (Fig. \ref{fig:Fig1}C).



\begin{figure*}[th!]
    \centering
    \includegraphics[width=18cm]{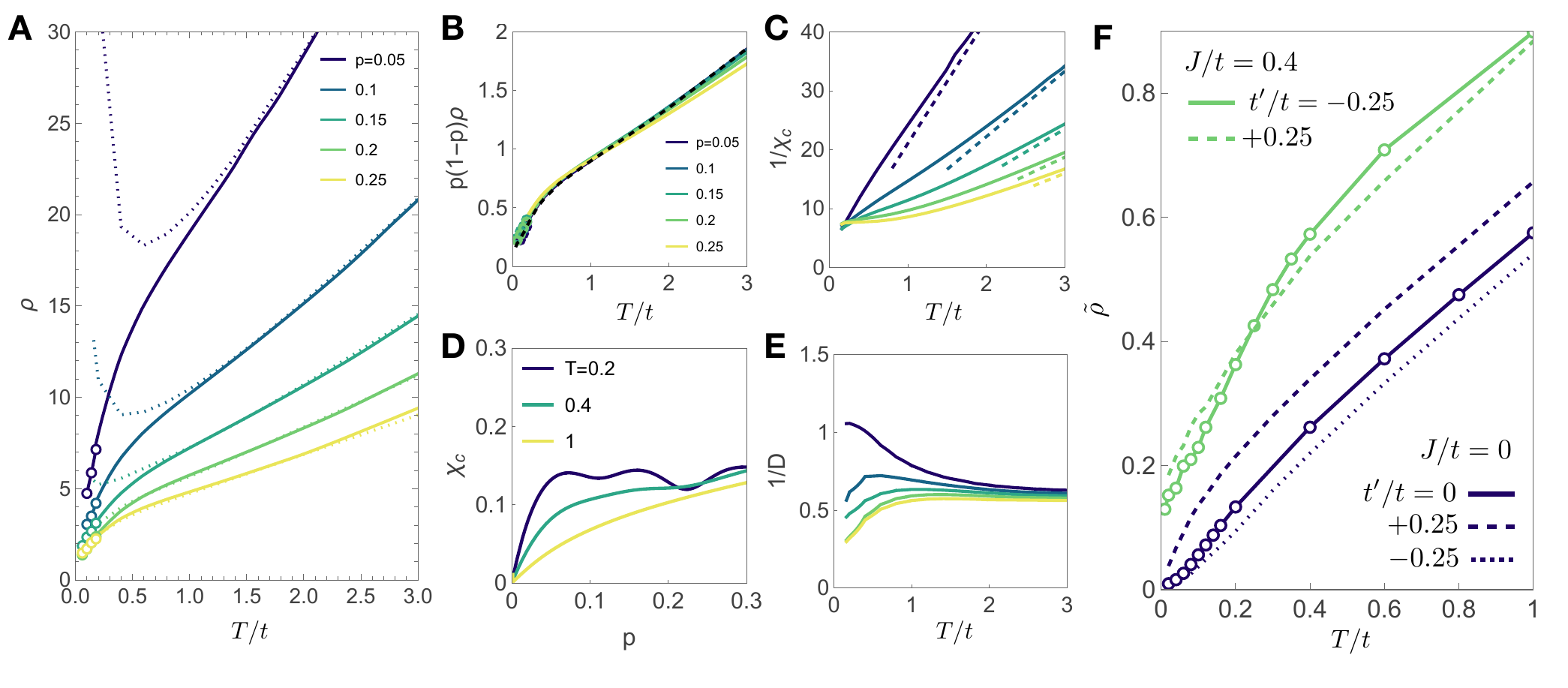}
    \caption{
   \textbf{Single-hole nature of strange metal transport.} 
   (\textbf{A}) Resistivity calculated from FTLM at different doping levels (labels) for $J=0.4$ and $t^\prime/t=-0.25$, both in the pristine (dashed) and magnetically frustrated case (solid line, $J^\prime/J=0.58$). Parameters are $N_s=18$, $M=800$, $N_\phi=170$. The open circles are LTLM results using $M=400$, $N_\phi=49$. The same labels also apply to panels B, C and E. 
   (\textbf{B}) Same data as in panel A, rescaled according to Eq. (\ref{eq:prho-scaling}).  
   (\textbf{C}) Temperature dependence of the inverse charge compressibility, in units of $t$. The dashed lines represent the classical limit $1/\chi_c=T/p/(1-p)$. At low $T$ and large $p$, the data saturate to a finite value (the inverse of the density of states at the Fermi level), as expected in a degenerate liquid.
   (\textbf{D}) Compressibility vs doping at different temperatures (labels).
   (\textbf{E}) Inverse diffusivity vs temperature, in units of $ta^2$, derived from the Nernst-Einstein relation (see text).
   (\textbf{F}) One-hole resistivity (inverse mobility) for different parameter choices (labels), calculated with $N_s=20$, $N_\phi=49$, $M=400$   (except $J=t^\prime=0$: $N_s=24$, $M=800$,  $N_\phi=105$).}
    \label{fig:Fig2}
\end{figure*}

\textit{Finite hole concentrations.---}
We now turn to finite carrier concentrations that are of direct  relevance to experiments, and set microscopic parameter values representative of the hole-doped cuprates: we take $t^\prime/t=-0.25$ frustrating the perfect nesting on the square lattice, and $J/t=0.4$, corresponding to $U/t=10$ in the Hubbard model. Because the AF order shown in the single-hole calculations persists up to finite concentrations  \cite{LugasPRB06,XuPRR22} (the critical doping is $p_c\simeq 0.14$, see supplement), we additionally include a next-nearest-neighbor exchange $J^\prime/J=0.58$ that frustrates the AF order \cite{Balents12,JiangKivelson}, revealing metallic behavior in a broader doping range. 

Fig. \ref{fig:Fig2}A shows the resistivity vs temperature calculated with FTLM at varying hole concentrations from underdoped ($p=0.05$) to overdoped ($p=0.25$). Since, owing to exponentially increasing stochastic fluctuations, the accuracy of the FTLM data degrades at low temperatures, for $T/t<0.2$ we resort to the LTLM, that is numerically more demanding but correctly recovers the $T=0$ limit  \cite{Aichhorn03} (open circles), using the protocol described in Ref. \cite{prelovsek}. 

At high temperatures, $T \gg J$, we recover the same Planckian behavior of slope $\alpha=1$ found in Fig. \ref{fig:Fig1}A, obtaining identical results with (solid lines) and without the frustrating term $J^\prime$ (dashed lines).  At lower temperatures, however, suppressing the AF insulating behavior reveals a second $T$-linear behavior occurring at $T\lesssim J$ \cite{DengPRL13,Kokalj,VucicevicPRL19,spin-Holstein,TremblayPNAS22}.
In this low-temperature regime the slope of the resistivity is $\approx 2$ times larger than at high temperatures. Assuming that the relaxation rate is still Planckian, the same Drude analysis carried out above would imply that the charge carriers are twice as heavy as the original holes, which is compatible with the formation of composite excitations dressed by the spin fluctuations \cite{TrugmanPRB88,DemlerPRX18,Batista25,Coleman89,SenthilPRL03,ChangRPP25,Chalopin-PNAS26,Punk-PRR20}.  

\textit{Scaling with carrier density.---}
Fig. \ref{fig:Fig2}B shows the same resistivity data, rescaled with the number of holes that would be available for transport if they obeyed classical statistics, $p(1-p)$. Surprisingly,  all curves corresponding to different concentrations fall on top of each other with very good accuracy, and coincide with the result $\tilde\rho$ obtained for a single hole (dashed line) (see also supplement). This allows us to write the following scaling form:
\begin{equation}
  \sigma(T,p)\simeq  p(1-p) \ \tilde\sigma (T)
  \label{eq:prho-scaling}
\end{equation}
with $\tilde\sigma=1/\tilde\rho$ the carrier mobility.
This suggests that in the explored  concentration and temperature range, the holes move essentially as independent carriers: their mobility 
is insensitive to both mutual interactions and exchange, being instead determined by the properties of the fluctuating background in which they are embedded. 

Since the quantity that obeys scaling is the resistivity itself, 
our efforts at understanding strange metals should start from here.
In support of this standpoint,
we stress that the Nernst-Einstein relation, $\sigma =1/\rho= \chi_c D$, that is usually applied to analyze the conductivity of  quantum fluids in terms of the product of the charge compressibility $\chi_c$ (counting the carriers available for transport) and their diffusivity $D$, is of little help here:
as originally pointed out in Ref. \cite{Kokalj}, and shown in Fig. \ref{fig:Fig2}C and E, these two quantities separately do not show any obvious $T$-linear dependence. We add that they do not exhibit obvious scaling behavior with density either, as shown in Fig. \ref{fig:Fig2}D, which compares positively with the recent experimental results of Ref. \cite{Kendrick-Greiner25}. 

Reframing our focus on the resistivity itself, we can then make some important remarks. 
First, Eq. (\ref{eq:prho-scaling}) should apply to classical non-degenerate liquids, yet it appears to hold down to $T\to 0$, in stark contradiction with Fermi liquids.  The observed behavior indicates instead that 
Fermi statistics play a minor role. 
This is 
indicative of the incoherent nature of the carriers,
resulting from the strong scattering that characterizes bad metals in general.
The observation that the carrier mobility is essentially independent of carrier concentration in a broad concentration range can rationalize why, in experiments, strange metal behavior is observed over wide ranges of doping, without requiring the existence of  extended quantum criticality 
\cite{ZaanenSciPost18,PhillipsScience22,Hussey23}. The scaling Eq. (\ref{eq:prho-scaling})  is observed experimentally in the cuprates to a very good approximation \cite{Ando,BarisicPNAS13}.

Second, the low-temperature linear behavior in Fig. \ref{fig:Fig2}A and B extrapolates to a finite $T=0$ intercept, here $\rho_0\simeq 0.15 \bar\rho$. When converted to physical units, this residual resistivity is of the order of $50 \mu \Omega$cm for $p=0.2$, comparable to the experimentally observed values \cite{Ando,Legros}. 
Strange metals, unlike normal metals, can sustain intrinsic elastic scattering down to $T\to 0$: a finite positive $\rho_0$ is suggestive of residual disorder persisting down to the lowest temperatures. In the present model this is due to abundant, slow spin fluctuations \cite{PrelovsekPRL04}, that persist down to $T\to 0$ without ordering, in agreement with recent experimental results on the cuprates \cite{CampbellNPhys25,VinogradPRB22} and being a plausible cause of the destruction of the Fermi liquid. 
In fact, our data in the doping range $0<p<0.25$ show no sign of normal metallic behavior
(with the exception of the Nagaoka state realized at $J=0$). It is an interesting question whether restoring charge fluctuations in the finite-$U$ Hubbard model can favor the recovery of the Fermi liquid state at low temperature.

Since according to Eq. (\ref{eq:prho-scaling}), these are predictive of the behavior at finite hole concentrations, we present in Fig. \ref{fig:Fig2}C various representative  resistivity curves  obtained from single-hole calculations of $\tilde \rho(T)$ on cluster sizes up to $N_s=24$.
The results at finite $J$ are dominated by magnetic fluctuations, and are only marginally affected by the kinetic frustration term, showing similar $T$-linear behavior and finite $\rho_0$ regardless of the sign of $t^\prime/t$ (green). The effect of kinetic frustration \cite{Shastry-NJP18,WangDevereaux25} is instead crucial at low $J$ (here $J=0$). There the hole-doped cases $t^\prime/t\le 0$ (blue solid and dotted) show an initial Fermi liquid behavior due to ferromagnetic ordering, which is instead absent in the electron-doped case $t^\prime/t > 0$ (blue dashed).



\begin{figure}[th]
    \centering
    \includegraphics[width=8.5cm]{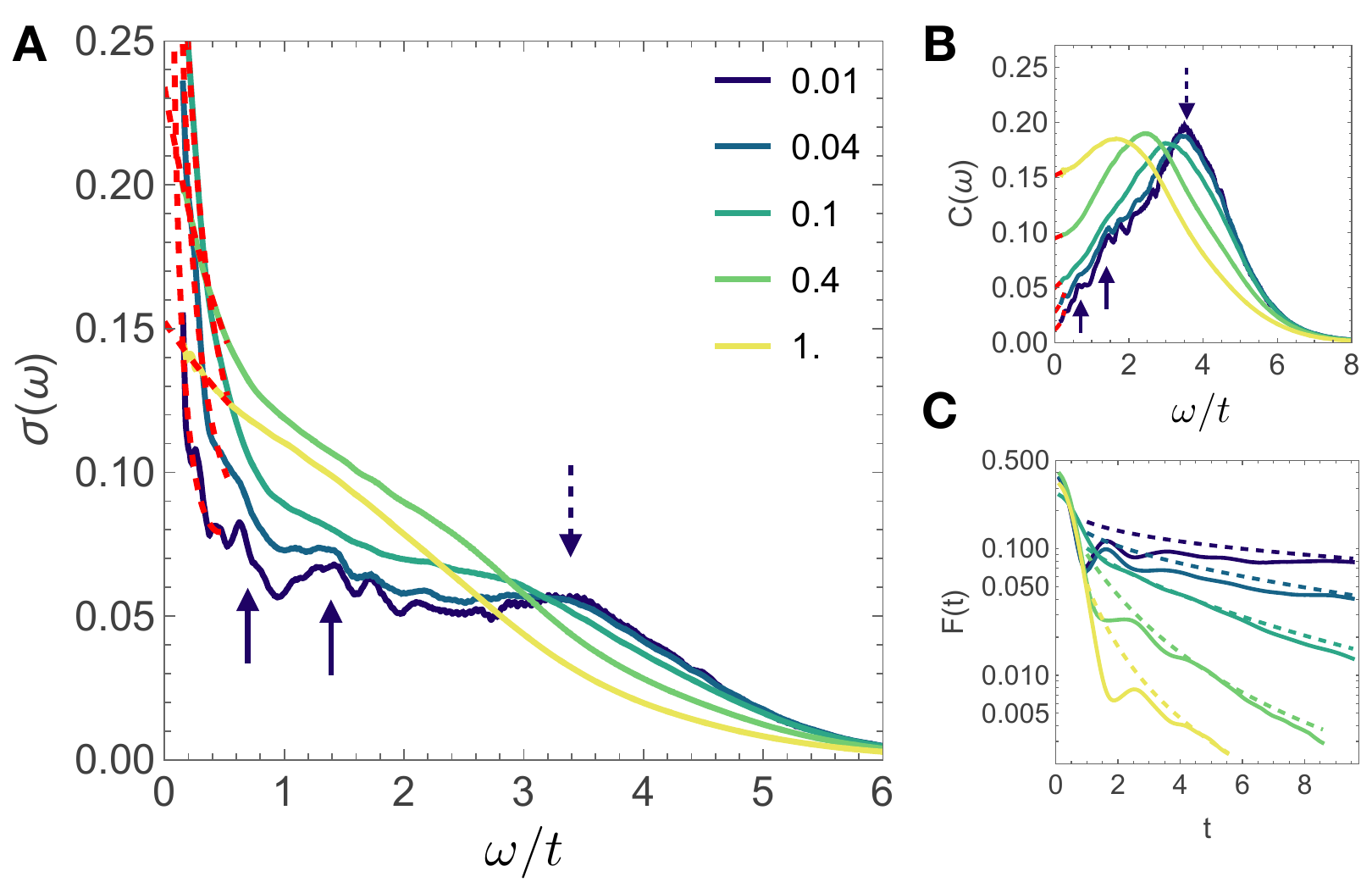}
    \caption{
   \textbf{Optical conductivity and time-resolved dynamics.} 
   (\textbf{A}) Optical conductivity calculated from LTLM at $J/t=0.4$, $t^\prime/t=-0.25$, $p=0.17$ and different temperatures (labels, the same apply to panels B and C). The dashed lines are two-parameter fits to the low-frequency part of the optical conductivity using Eq. (\ref{eq:Kubo}) with $C(\omega)=C_0/(1+C_1 \omega)$.
   The arrows mark the main magnetic (spin-flip) excitation at $\omega_J\simeq 1.7J$  and its first overtone (solid), in agreement with $T=0$ results \cite{Poilblanc,Poilblanc-opt91,VojtaEPL97}, as well as the (holon) charge excitation  continuum (dashed) \cite{Bang-Kotliar93}. 
   (\textbf{B}) Current-current correlation function $C(\omega)$ at the same temperatures. 
   (\textbf{C}) Fourier transform of the optical conductivity. Time $\mathrm{t}$ is measured in units of the inverse of the hopping integral, $\hbar/t$. The dashed lines represent the non-exponential long-time relaxation Eq. (\ref{eq:log}). 
      }
    \label{fig:Fig3}
\end{figure}

\textit{The strange metal in the frequency domain, and the origin of Planckian relaxation.---}
Up to now we have explored how strange  metallicity develops vs temperature $T$ (linearity, Eq. \ref{Drude-analysis}) and density $p$ (scaling, Eq. \ref{eq:prho-scaling}). We now show that a much deeper understanding of strange metals can be gained by exploring the frequency domain, $\omega$.

Fig. \ref{fig:Fig3}A shows the temperature evolution of the optical conductivity at a doping $p=0.17$, for $J/t=0.4$, $J^\prime/J=0.58$ and $t^\prime/t=-0.25$. Starting from  $T/t=1$, within the high temperature regime where the carriers are undressed holes (cf. Fig \ref{fig:Fig2}A), we can already see that the optical absorption does not follow the 
characteristic Lorentzian shape corresponding to Drude theory, 
$\sigma_D(\omega)=\sigma_D/(1+\omega^2\tau_{\mathrm{tr}}^2)$: the decay is instead linear in frequency, compatible with the cusp-like behavior $\sigma(\omega)=\sigma(0)-A | \omega | $ that was rigorously established in the ultra-high temperature limit in two dimensions \cite{HusePRB06}.

Even more interesting is the behavior observed at low temperature. For $T\lesssim J$, i.e. where the resistivity indicates dressing of the holes into composite particles, the optical absorption starts exhibiting a more complex structure, with a high-energy continuum located at $0.5t\lesssim \omega \lesssim 3.5t$  and a narrower contribution developing at lower frequencies. The features of the high-energy continuum reflect the internal structure of the composite carriers, which comprises both magnetic \cite{Poilblanc-opt91,Poilblanc,VojtaEPL97} and charge excitations \cite{Bang-Kotliar93}, indicated respectively by full and dashed arrows. The $\omega\approx 0$ contribution represents instead their long-distance motion: this part is crucial, as it is the one that determines the strange metal behavior observed in the resistivity.

The width of the low-frequency peak scales proportionally to $T$, 
an observation that, at first sight, seems compatible with a relaxation time of the form of Eq. (\ref{Drude-analysis}).
The high spectral resolution of our data, however, allows us to discriminate the precise shape of this low-energy absorption peak, which is conclusively not compatible with the Drude formula: at low temperature
the decay appears to be  $1/\omega$ instead of $1/\omega^2$ predicted by Drude theory \cite{Prelovsek-PRL95,Prelovsek-EPL01,minimal-PRL26,RiceZhangPRB89}.

Since the Drude absorption shape follows from the same hypothesis of exponential relaxation in time that determines the d.c. conductivity $\rho_D=m/(ne^2\tau_{\mathrm{tr}})$ \cite{GrunerDresselBook02}, this inconsistency 
undermines the applicability of the Drude analysis altogether. The optical conductivity should instead be understood based on the
Kubo formula: 
\begin{equation}
    \sigma(\omega)=\frac{1-e^{-\hbar\omega/k_BT}}{\hbar\omega} C(\omega),
    \label{eq:Kubo}
\end{equation}
which is an exact equation relating 
it to the current-current correlation function $C(\omega)=\mathrm{Re}\int dt e^{i\omega \mathrm{t}} \langle J(\mathrm{t}) J(0) \rangle$. As pointed out in Refs. \cite{RiceZhangPRB89,Prelovsek-PRL95,Prelovsek-EPL01,minimal-PRL26}, a featureless $C(\omega)\approx C_0$ on the relevant frequency scale $\omega \sim 1/\tau_{\mathrm{Planck}}=k_BT/\hbar$--- which is a defining feature of bad metals \cite{Gunnarsson,Hussey} ---
implies that the shape of the optical absorption is mostly determined by the detailed balance prefactor $(1-e^{-\hbar\omega/k_BT})/\omega$, leading to  the observed ``stretched" $1/\omega$ decay, with an effective width set by $1/\tau_{\mathrm{Planck}}$ itself.
A $T$-linear resistivity  as well as  the marginal Fermi liquid phenomenology  \cite{VarmaMFL} revealed in optical experiments, including the transport-derived  apparent
rate Eq. (\ref{Drude-analysis}), immediately follow from assuming a $T$-independent $C_0$ \cite{Prelovsek-PRL95,Prelovsek-EPL01,minimal-PRL26}. 
We add that at high $T$, expanding Eq. (\ref{eq:Kubo})  with $C(\omega)=C_0$ yields $\sigma(\omega)=(C_0/k_BT)(1-|\hbar \omega/k_BT|/2)$, which entails the cusp-like behavior observed in Fig. \ref{fig:Fig3}A. Kramers-Kroing transforming the same expression at low $T$  leads to an apparent $|\sigma|\sim 1/\omega^{0.8}$ in agreement with the exponent reported in Ref. \cite{MichonNatComm23}  (see supplement for the resulting analytical expression for the complex conductivity) and a characteristic bad-metal reflectivity lacking a sharp plasma edge \cite{BasovRMP,vanHeumen22}.  

To verify this entire scenario, we report in Fig. \ref{fig:Fig3}B the current-current correlation function  corresponding to $\sigma(\omega)$ of Fig. \ref{fig:Fig3}A. As we can see clearly,  $C(\omega)$ 
does not contain any intrinsic narrow features occurring on the Planckian scale $\omega \lesssim 1/\tau_{\mathrm{Planck}}$. Moreover, upon lowering the temperature the entire function uniformly  converges to  a well-defined zero-temperature limit. 
These two observations imply that the narrow peak emerging at low temperature in Fig. \ref{fig:Fig3}A, as well as the $T$-linear dependence of the resistivity itself, are both direct consequences of the prefactor in Eq. (\ref{eq:Kubo}). The apparent scattering rate $1/\tau_{\mathrm{\mathrm{tr}}}=k_B T/\hbar$, inferred experimentally from transport and optical measurements, follows naturally from the scaling factor $\hbar\omega/k_BT$ that governs detailed balance.  This scenario also implies that the $w=0$ peak in the optical conductivity should keep narrowing upon lowering the temperature even when the overall spectral weight is suppressed by pseudogap phenomena, as observed in the cuprates \cite{HwangTimusk-JPCM07,vanHeumen22}.

\textit{Planckian relaxation in the time domain.---}
We finally adress a question that has been taking shape throughout this work:
If not exponential, what is the distinctive time-dependence of Planckian relaxation?
Fig. \ref{fig:Fig3}C illustrates  the Fourier transform  of the optical conductivity, $F(\mathrm{t})=\int \sigma(\omega) \cos(\omega \mathrm{t}) d\omega$ that, in the classical case, embodies the decay of the carrier velocity correlations in the time domain. The Drude model is built by assuming simple exponential relaxation, $F(\mathrm{t})\propto e^{-\mathrm{t}/\tau_{\mathrm{tr}}}$ \cite{GrunerDresselBook02}. The data, however, show a more complex behavior.

The bad metal character of the system causes very fast initial relaxation at short times, on the timescale of the hopping itself: carrier coherence is lost at each hop, in line with theoretical proposals that do not rely on the existence of quasiparticles \cite{PatelPRX25,PatelScience23}. The initial decay features damped oscillations, with a frequency given by the holon peak in Fig. \ref{fig:Fig3}B: this is the same local (momentum independent) incoherent excitation seen by charge probes \cite{MitranoPNAS18,Arpaia}. 
The data suggest however the emergence of a second, much slower type of relaxation
that sets in at long times \cite{minimal-PRL26}, being responsible for both the strange metal resistivity and the non-Drude shape of the optical conductivity. 
This long-time decay can again be understood by setting $C(\omega)=C_0$ in Eq. (\ref{eq:Kubo}).  Performing the frequency-integral yields
\begin{equation}
    F(\mathrm{t})=\frac{C_0}{2} \ \log \frac{1}{1+(\mathrm{t}/\tau_{\mathrm{Planck}})^2},
    \label{eq:log}
\end{equation}
corresponding to the dashed lines in Fig. \ref{fig:Fig3}C. The distinctive feature is the logarithm, that stems from the singular $1/\omega$ behavior of $\sigma(\omega)$, cut off by the Planckian scale. 

This example illustrates once more how in bad metals, where the quantum prefactor in the relevant response function becomes dominant, the analysis of $\sigma(\omega)$ can give the impression of an additional relaxation channel. The actual relaxation of the current, however, is given by the Fourier transform of $C(\omega)$, which only exhibits one, fast component.

\textit{Concluding remarks.---}
The numerical results presented throughout this work provide compelling evidence of strange metal behavior arising in a strongly correlated system when the charge carriers are coupled to --- and move within --- a strongly fluctuating magnetic environment. 
Beyond the specifics of the model studied here, our findings point
to a general explanation of Planckian relaxation and call for a change of paradigm:
in bad metals, 
where the decay of the electrical current is very fast,
Planckian behavior is not entailed in the scattering properties of the charge carriers, but arises instead from the very quantum statistical nature of their response to electric fields.

\bibliography{PlanckiantJ} 

\section*{Methods}

The finite-temperature Lanczos method (FTLM)  \cite{prelovsek} provides an exact high-temperature expansion for static and dynamic correlation functions, making efficient use of the idea of typicality \cite{Schnack} to map the problem onto a reduced number of pseudo-eigenstates. The size of the resulting Krylov space, $M$, is set by the total number of performed Lanczos iterations, which is much smaller than the size of the Hilbert space itself. In practice we use values in the range $400\le M\le 2400$. For comparison, in the largest system studied here ($N_s=20$ sites, $N_p=15$ particles, spin sector $S_z=0$) the Hilbert space is composed of $\approx 6\times 10^7$ states. 

To circumvent the exponential increase of stochastic fluctuations characterizing the FTLM in the low temperature limit, when needed we use the the low-temperature generalization of the method, LTLM, that is numerically more demanding but correctly recovers the $T=0$ limit  \cite{Aichhorn03}. 

Our numerical calculations are performed on clusters of up to $N_s=24$ sites in the case of a single hole, and up to $N_s=20$ at finite hole concentrations, restricting to the lowest $S_z$  spin sector. We additionally use twisted boundary condition (TBC) averaging to augment the effective number of sites. This increases the number of optical transitions, hence increasing the spectral resolution. 

The optical conductivity, $\sigma(\omega)$, is evaluated through the Kubo formula Eq. \ref{eq:Kubo}, where the current-current correlation function $C(\omega)$ is calculated by summing over all of the $M^2$ transitions between different pseudo-eigenstates. Transport properties are evaluated by taking the limit $\omega\to0$ of the optical conductivity. This is achieved by extrapolating the current-current correlation function as $C(\omega)=C_0/(1+C_1\omega)$ (red dashed lines in Fig. \ref{fig:Fig3}A,B) and then using Eq. (\ref{eq:Kubo}) to obtain $\rho=T/C_0$.  This constitutes an improved and physically motivated version of the protocol originally devised in Ref. \cite{RammalPRL24} to deal with the spurious terms introduced at $\omega\approx 0$ by the
fictitious magnetic fluxes involved in the TBC averaging procedure. 

When appiled to the t-J model, the present methodology allows us to obtain reliable results for charge transport down to the experimentally relevant window, reaching $T/t=0.01$. The high spectral resolution $\Delta \omega=0.002t$ gives access to the low-frequency/long-time charge dynamics, which is crucial for the phenomena revealed in this work.

For the finite-density calculations of Fig.\ref{fig:Fig2} and Fig. \ref{fig:Fig3}  we work in the grand canonical ensemble, using the prescription of Ref. \cite{Gros1992} for TBC averaging, that we generalize here to finite temperatures.  Calculations targeting the properties of individual holes (Fig. 1A,B,C and Fig. 2F) are performed in the canonical ensemble, using the prescription of Ref. \cite{Poilblanc}, whose validity is demonstrated in the supplementary material.


\section*{Acknowledgments}
S.F. acknowledges 
useful discussions with  K. Behnia, M. Dressel, A. Georges, T. Giamarchi,  D. Golez, E. Heller, M.-H. Julien, L. K. Kendrick, J. Kokalj, D. Le Boeuf, D. van der Marel, L. de' Medici, A. J. Millis, J. Mravlje, P. Prelov\v{s}ek, R. Queiroz, L. Rademaker, H. Rammal, Z.-X. Shen and C. Swickard. S.F. is particularly grateful 
to  M.-H. Julien for suggesting the study of the frustrated quantum paramagnet, to M.-B. Lepetit for sharing her calculation resources and to the Biblioteca de Catalunya and Jo\v{z}ef Stefan Institute, where the initial version of this manuscript was laid out.
S.C. acknowledges useful discussions with S. Caprara. S.F. received support  from the French Agence Nationale de
la Recherche (ANR) under Grant Reference No. ANR-25-CE30-2817 and S.C. from NextGenerationEU National Innovation Ecosystem Grant No. ECS00000041- VITALITY-CUP E13C22001060006.
This project was provided with computing HPC and storage resources by GENCI at CINES thanks to the grants 2025-AD010916992 and 2026-AD010917651 on the supercomputer Adastra's GENOA partition.

\textit*{Author contributions:}
S.F. had the initial idea, led the project, devised the methodology, wrote the initial FTLM code, the LTLM code and the manuscript.  A.R. wrote the initial FTLM code and helped with the intepretation of the results. I.D. improved and rewrote both codes and helped with the intepretation of the results. SC helped with the intepretation of the results.


\end{document}